\documentstyle[12pt,epsf,psfig]{article}
\begin{document}
\baselineskip 0.6cm
\newcommand{\gsim}{ \mathop{}_{\textstyle \sim}^{\textstyle >} }
\newcommand{\lsim}{ \mathop{}_{\textstyle \sim}^{\textstyle <} }
\newcommand{\vev}[1]{ \langle {#1} \rangle }
\newcommand{\EV}{ {\rm eV} }
\newcommand{\KEV}{ {\rm ~keV} }
\newcommand{\MEV}{ {\rm ~MeV} }
\newcommand{\GEV}{ {\rm ~GeV} }
\newcommand{\TEV}{ {\rm ~TeV} }
\newcommand{\eps}{\varepsilon}
\newcommand{\barr}[1]{ \overline{{#1}} }
\newcommand{\del}{\partial}
\newcommand{\nn}{\nonumber}
\newcommand{\ra}{\rightarrow}
\newcommand{\bino}{\tilde{\chi}}
\def\tr{\mathop{\rm tr}\nolimits}
\def\Tr{\mathop{\rm Tr}\nolimits}
\def\Re{\mathop{\rm Re}\nolimits}
\def\Im{\mathop{\rm Im}\nolimits}
\setcounter{footnote}{0}

\begin{titlepage}

\begin{flushright}
CERN-TH/2002-124\\
UT-02-37
\end{flushright}

\vskip 3cm
\begin{center}
{\Large \bf A Solution to the Coincidence Puzzle \\
of $\Omega_{B}$ and $\Omega_{DM}$}
\vskip 2.4cm

\center{Masaaki~Fujii$^{a,b}$, and T.~Yanagida$^{a,b,}$}\footnote{On 
leave from University of Tokyo.}\\
$^{a}${\it{CERN Theory Division, CH-1211 Geneva 23, Switzerland }}\\
$^{b}${\it{Department of Physics, University of Tokyo, Tokyo 113-0033,
Japan}}\\
\vskip 2cm

\abstract{ We show that a class of Affleck--Dine baryogenesis directly
  relates the observed mass density of baryons, $\Omega_{B}$, to
  that of dark matter, $\Omega_{DM}$.  In this scenario, the ratio
  of baryon to dark matter mass density is solely determined by the
  low energy parameters, except for an ${\cal O}(0.1)$ effective
  CP-violating phase.
  We find that $\Omega_{B}/\Omega_{DM}={\cal O}(0.1)$
with  reasonable parameters, which lies surprisingly just in
the range of observation.
 This scenario is totally free from the cosmological gravitino
 problem, and independent of the detailed history of the Universe as
 long as it satisfies quite weak constraints.  }

\end{center}
\end{titlepage}

\renewcommand{\thefootnote}{\arabic{footnote}}
\setcounter{footnote}{0}
%
%
%
\section{Introduction}
The searches for both the origin of the observed baryon  asymmetry and 
that of the dark matter  in the
present Universe have been attractive subjects for many physicists.
The great success of the Big Bang nucleosynthesis (BBN) predicts an
abundance of the baryon asymmetry in the present Universe as
$\Omega_{B}h^2\simeq 0.02$.
Here, $h$ is the present Hubble parameter in units of $100\;{\rm
km}\;{\rm sec}^{-1}\;{\rm Mpc}^{-1}$, and $\Omega_{B}\equiv \rho_{B}/\rho_{c}$.
($\rho_{B}$ and $\rho_{c}$ are the energy density of the baryon and the
critical energy density in the present Universe.)
As for the dark matter, various observations of the dynamics of
galaxies and their clusters, as well as theoretical analyses of the
large-scale structure formation 
indicate that its cosmological abundance lies in the range, $\Omega_{\rm
DM}h^2\simeq 0.1$--$0.2$.

There exist a number of scenarios of ``baryogenesis'', which
can generate the required baryon asymmetry.
Also for dark matter, several particles have been proposed as
candidates, such as axion, neutrino and, among others, the lightest
supersymmetric (SUSY) particle (LSP) (especially the lightest
neutralino, $\chi$).

However, there remains one prominent problem.  The matter density of
baryons is roughly ${\cal O}(10$--$20)$\% of that of dark matter, {\it
  i.e.} $\Omega_{B}/\Omega_{DM}\simeq 0.1$--$0.2$.  Why are they so
close to each other?  Most of the models of baryogenesis use baryon
$B$- and/or lepton $L$- number-violating operators in high energy
physics.  On the other hand, the abundance of dark matter does not
seem to be at all related with such high energy physics relevant to
baryogenesis.  In the standard SUSY dark matter scenario, for example,
the mass density of the dark matter is solely determined by the weak
scale properties of the neutralino LSP, such as the mass and the
annihilation cross section.  As long as dark matter is composed of
the thermal relics of the LSPs, it is completely independent of other
details in the history of the Universe.  It must be very exciting if
we can find a simple mechanism to directly relate the two independent
physical parameters, the baryon asymmetry and the dark matter density.

One interesting solution to this puzzle was proposed by using the
Affleck--Dine (AD) baryogenesis~\cite{AD} in the context of the
minimal supergravity (mSUGRA) scenario~\cite{B-ball-org}.~\footnote{
  The possibility to explain both the baryon asymmetry and  dark
  matter by a single mechanism was first considered in the context of
  gauge-mediated SUSY breaking scenarios~\cite{K-S}.  }  In this
scenario, non-topological solitons called Q-balls~\cite{Coleman} play
the role of the common source of the baryon asymmetry and the LSP.  In
the AD baryogenesis, a complex scalar field $\phi$, which is a linear
combination of squark fields, obtains a large expectation value along
one of the flat directions in the scalar potential during inflation.
The subsequent coherent oscillation of the $\phi$ field obtains a
phase rotational motion and begins to carry non-zero baryon number in
the presence of $B$-violating operators.  A crucial point is that this
coherent oscillation of the $\phi$ field is unstable, with spatial
perturbations and fragments into the Q-balls after dozens of
oscillations.  Recent detailed lattice simulations have revealed that
almost all the baryon asymmetry initially carried by the coherent
$\phi$ field is absorbed into Q-balls~\cite{KK-1,KK-2}.

The large amplitude of the $\phi$ field inside the Q-ball protects it
from being thermalized.  The typical decay temperature of Q-balls is
given by $T_{d}\lsim (\mbox{a few})\GEV$, which is well below the
freeze-out temperature of the LSP. As a result, the LSPs produced by
the Q-ball decay are not thermalized and retain the initial abundance.
Therefore, the mass density of the non-thermal neutralino dark matter
$\Omega_{\chi}$ is directly connected to that of
baryons~\cite{B-ball-org}:
\begin{equation}
\Omega_{\chi}\simeq \left(\frac{m_{\chi}}{m_{p}}\right)\left(
\frac{n_{B}}{n_{\phi}}\right)^{-1}\times\Omega_{B}\;,
\label{relic-neutralino}
\end{equation}
where $m_{\chi}$ and $m_{p}$ are the LSP and 
nucleon
masses, respectively; $n_{B}$ and $n_{\phi}$ are the 
baryon and $\phi$ field number densities.
One can easily understand this relation by considering that
the decay of a single $\phi$ field produces $\gsim 1$ neutralino LSP in
the final state.
Here, the ratio of $n_{B}$ to
$n_{\phi}$ is fixed when the $\phi$ field starts coherently oscillating and
remains constant until it eventually decays.
Under the assumption of $R$-parity conservation, the possible maximum
value of this ratio is $(n_{B}/n_{\phi})=1/3$.
The typical value of this ratio, which naturally appears in models of
AD baryogenesis in the mSUGRA scenario is
$n_{B}/n_{\phi}\lsim 0.1$.

Unfortunately, this simple model is not cosmologically viable.
We need an extremely light bino $m_{\chi}\lsim 1\GEV$ to explain the 
required mass density of dark matter.
This fact indicates that the formation of a Q-ball is a serious obstacle
for the AD baryogenesis rather than a solution
to the $\Omega_{B}$--$\Omega_{DM}$ coincidence puzzle.~\footnote{
In the original work, the baryon absorption into the produced Q-balls
was assumed not to be so efficient, which conflicts with the
detailed lattice simulations~\cite{KK-1,KK-2}.}

There are several ways to reconcile this situation.  The obvious one
is to make the Q-balls small enough that they can evaporate well
above the freeze-out temperature of the LSP.  This can be done by
gauging the U(1)$_{B-L}$ symmetry~\cite{ADwithBL} or assuming a
relatively strong three-point coupling of the inflaton to the
gluino~\cite{HC}.  In both cases, it is clear that the relation
between $\Omega_{B}$ and $\Omega_{\chi}$ is lost completely.  
Another possibility is to adopt the LSP with a large annihilation
cross section, such as higgsino $\widetilde{H}$ or wino
$\widetilde{W}$.  In Refs.~\cite{higgsino-wino1, higgsino-wino2}, it is
shown that the subsequent pair annihilations of these LSPs can
naturally lead to the desired mass density of dark
matter.~\footnote{The LSP is not necessarily a pure $\widetilde{H}$ or
  a pure $\widetilde{W}$. A significant mixing of the bino component
  in the LSP is quite possible. In the case of a large ${\rm
    tan}\beta$, the acceptable mass density of dark matter can be
  obtained even with the LSP whose dominant component is a bino.}
However, in this case, one expects that the resultant mass density of
baryons and that of dark matter are completely independent of each
other, since the number of produced LSPs should be drastically reduced
via annihilation processes so as not to overclose the Universe.  However,
astonishingly, this is not always the case.

In this letter, we will show that a class of AD baryogenesis
scenarios
allow us to  connect the two crucial quantities in
our Universe.
Our final goal is to show the following relation between the mass
density of baryons and that of dark matter:~\footnote{As we will see,
when the mass of the gravitino $m_{3/2}$ is larger than $m_{\phi}$,
$m_{\phi}$ in Eq.~(\ref{therelation}) should be replaced by $m_{3/2}$.}
\begin{equation}
\frac{\Omega_{B}}{\Omega_{\chi}}\simeq 10^{3{\mbox{--}}4}\left(
\frac{m_{\phi}^2}{\vev{\sigma v}_{\chi}^{-1}}\right)
\left(\frac{m_{p}}{m_{\chi}}\right)\delta_{\rm eff}\;,
\label{therelation}
\end{equation}
where $m_{\phi}$ is the mass of the flat direction field $\phi$,
which is approximately given by the squark mass;
$\vev{\sigma v}_{\chi}$ is the $s$-wave component of the annihilation
cross section of the LSP; $\delta_{\rm eff}={\cal{O}}(0.1)$ is the effective
CP-violating phase of the $\phi$ field. A numerical factor in the
right-hand side depends on the details of the produced Q-balls, but
basically they are calculable.
If we take the typical annihilation cross section of the
$\widetilde{H}$- or $\widetilde{W}$-like LSP, $\vev{\sigma
v}_{\chi}\sim 10^{-(7{\mbox{--}}8)}\GEV^{-2}$, one can easily see
that this is just the desired relation.
In the remainder of this paper, after a
brief review of the AD baryogenesis,
we will derive this interesting relation
and discuss the conditions for it to hold.

\section{The model}
Our model of the AD baryogenesis is basically the same as
the one presented in Section II B of Ref.~\cite{higgsino-wino2}.
For self-consistency, we briefly review the mechanism
with particular attention to the late-time decays of Q-balls.
The derivation of the relation  Eq.~(\ref{therelation})
and some conditions for it to hold will be discussed in subsequent 
sections.

We consider the situation that there exist some chiral symmetries, such
as $R$-symmetry, which forbid non-renormalizable operators in the
superpotential that lift the relevant flat directions for baryogenesis.
$B$-violating operators needed for AD baryogenesis are
supplied in the K\"ahler potential. We take the following operators,
for example,
which are consistent with the $R$-symmetry:
\begin{equation}
\delta{\cal L}=\int d^{4}\theta
\left(
\lambda_{1}\frac{I^{\dag}I}{M_{*}^4}Q\bar{U}^{\dag}\bar{D}^{\dag}L
+\lambda_{2}\frac{Z^{\dag}Z}{M_{*}^4}Q\bar{U}^{\dag}\bar{D}^{\dag}L+{\rm h.c.}
\right)
\;,
\label{Kahler}
\end{equation}
where $M_{*}=2.4\times 10^{18}\GEV$ is the reduced Planck scale, and
$\lambda_{1},\lambda_{2}$ are the coupling constants of the order of $1$,
which are generally complex numbers. Here,
$Q$, $\bar{U}$, $\bar{D}$, $L$, $I$ and $Z$ denote superfields (and their
scalar components) of
left-handed quark doublets, right-handed up-type and down-type quarks,
left-handed lepton doublets, inflaton and the one relevant to SUSY
breaking in the true vacuum,
respectively.~\footnote{The $B$-violating operator $\propto
QQ\bar{U}^{\dag}\bar{E}^{\dag}$ can also be applied to the following arguments.}
These interactions induce the following terms in the scalar potential:
\begin{equation}
 \delta V=-\left(
\lambda_{1}\frac{3 H^2}{M_{*}^2}Q\bar{U}^{\dag}\bar{D}^{\dag}L
+\lambda_{2}\frac{3 m_{3/2}^2}{M_{*}^2}Q\bar{U}^{\dag}\bar{D}^{\dag}L+{\rm h.c.}
\right)\;,
\label{potential1}
\end{equation}
where $m_{3/2}$ is the mass of the gravitino, and $H$
is the Hubble parameter of the expanding Universe.

If the expectation value $\vev{Q\bar{U}^{\dag}\bar{D}^{\dag}L}$ is
large enough after the inflation, the terms in Eq.~(\ref{potential1})
give the phase rotational motion to this scalar condensate and
generate net baryon asymmetry.  In fact, in the scalar potential of
the minimal supersymmetric standard model (MSSM), there are many $D$-
and $F$-flat directions~\cite{flat-directions} that give a non-zero
expectation value of $\vev{Q\bar{U}^{\dag}\bar{D}^{\dag}L}$.  In the
remaining discussion, we adopt the flat direction labelled by a linear
combination of the monomials of the chiral superfields:
$\bar{U}\bar{D}\bar{D},\;Q\bar{D}L$.  An explicit parametrization of
this direction, as suggested in the original 
work~\cite{AD}, is given by
\begin{eqnarray}
&&\qquad \qquad\quad\quad Q^{(1)}=\left(
\begin{array}{ccc}
\phi_{1}&0&0\\
0&0&0\end{array}
\right),\;
L^{(1)}=\left(
\begin{array}{cc}
0\\\phi_{1}\end{array}
\right)\;,\nonumber\\
&&\bar{D}^{(2)}=\left(
\begin{array}{ccc}
\phi_{3}&0&0\end{array}
\right),\;
\bar{U}^{(1)}=\left(
\begin{array}{ccc}
0&\phi_{2}&0\end{array}
\right),\;
\bar{D}^{(1)}=\left(
\begin{array}{ccc}
0&0&\phi_{2}
\end{array}
\right)\;,\nonumber\\
&&\qquad\qquad\qquad\qquad\qquad |\phi_{3}|^2=|\phi_{1}|^2+|\phi_{2}|^2\;,
\label{flat-parametrization}
\end{eqnarray}
where superscripts $(1)$--$(3)$ 
denote the generations of the chiral superfields.
Here, the column and low vectors denote the isospin of the
SU(2)$_{\rm L}$ and the colour of the SU(3)$_{\rm C}$ gauge groups of
the MSSM, respectively.
Hereafter, we use this simple parametrization.
Although
there are many other parametrizations of the
flat direction,
the main arguments in the following discussion do not
change much.

To give large expectation values to the flat-direction
fields, we assume the
four-point couplings of these fields to the inflaton in the K\"ahler
potential~\cite{DRT}:
\begin{eqnarray}
\delta K =\frac{I^{\dag}I}{M_{*}^2}\left(b_{1}\phi_{1}^{\dag}\phi_{1}
+b_{2}\phi_{2}^{\dag}\phi_{2}+b_{3}\phi_{3}^{\dag}\phi_{3}
\right)\;,
\label{K-negative}
\end{eqnarray}
where $b_{i}$'s are real coupling constants of the order of $1$.
Here we use the same symbols for the chiral superfields as for the
corresponding scalar components in Eq.~(\ref{flat-parametrization}).
By calculating the scalar potential of the supergravity, we can find
that the following terms are induced at leading order:
\begin{equation}
\delta V=3\left[1-(b_{1}+b_{3})\right]H^2|\phi_{1}|^2+
3\left[1-(b_{2}+b_{3})\right]H^2|\phi_{2}|^2\;.
\end{equation}
Therefore, if both $(b_{1}+b_{3})$ and $(b_{2}+b_{3})$ are somewhat
larger than $1$, $\phi_{1}$
and $\phi_{2}$ (and so $\phi_{3}$) obtain large expectation values
during inflation. Furthermore, if $(b_{1}+b_{3})\approx (b_{2}+b_{3})$,
we can naturally expect that $|\phi_{1}|\approx|\phi_{2}|\; (\equiv
|\phi|)$.
We assume that this is the case in the remainder of this paper.

Now, we can approximately write down the relevant scalar potential of the $\phi$
field as follows:
\begin{equation}
 V=(m_{\phi}^2-c_{H}H^2)|\phi|^2+\frac{H^2}{4M_{*}^2}(a_{H}\phi^4+{\rm h.c.})
+\frac{m_{3/2}^2}{4 M_{*}^2}(a_{m}\phi^4+{\rm h.c.})+\ldots\;,
\label{potential-phi}
\end{equation}
where the ellipsis denotes the higher-order terms coming from the
K\"ahler potential; $a_{H}$, $a_{m}$ are complex coupling constants;
$c_{H}={\cal{O}}(1)$ is a real coefficient; $m_{\phi}$ denotes the soft
SUSY-breaking mass
of the $\phi$ field, which is roughly given by the average value of the
soft SUSY-breaking masses of the $Q$, $\bar{U}$, $\bar{D}$ and $L$.
Here, we have neglected the terms induced by the interactions with
thermal backgrounds, which will be justified later.

We are, now, at the point where we want to discuss the evolution of
the $\phi$ field.  During inflation, the large negative Hubble-induced
mass term causes an instability around the origin, and the $\phi$ field
develops a large expectation value.  The amplitude of the $\phi$ field
just after the inflation, $|\phi_{0}|$, is determined by the balance
point between the Hubble-induced mass term and non-renormalizable
operators in the K\"ahler potential.  In the following discussion, we
treat it as a free parameter and assume $|\phi|_{0}\lsim M_{*}$.
(Actually, there exists an interesting method to fix $|\phi|_{0}$ below
the Planck scale. This can be done by gauging the U(1)$_{B-L}$
symmetry.  In this case, we can fix the amplitude of the $\phi$ field
at the $B-L$ breaking scale, $|\phi|_{0}\simeq
v_{B-L}$.~\footnote{Although the operator conserves the $B-L$
  symmetry, the flat directions carry non-zero $B-L$ charges with the
  same sign; then, they can be lifted at the $B-L$ breaking scale
  by the potential induced by the U(1)$_{B-L}$
  $D$-term~\cite{ADwithBL}.}  The details on this point are discussed
in Ref.~\cite{ADwithBL}.)

After the end of inflation, the amplitude of the Hubble parameter
gradually decreases. When $m_{\phi}$ exceeds the Hubble parameter, the
$\phi$ field starts coherent oscillation around the origin.  At this
time $H=H_{\rm osc}\simeq m_{\phi}$, a huge baryon asymmetry is produced
because of the second and third operators in
Eq.~(\ref{potential-phi}).  As long as $|\phi|_{0}\lsim M_{*}$, the
curvature along the phase direction is smaller than the Hubble parameter
during the inflation.  Therefore, the initial phase of the $\phi$ field
is generally displaced from the bottom of the valley of the scalar
potential.
This is the reason why the second and third operators give the phase
rotational
motion to the $\phi$ field.

It is not difficult to
estimate the ratio of baryon to $\phi$-number density, which is fixed at
$H=H_{\rm osc}$, as
\begin{equation}
 \left(\frac{n_{B}}{n_{\phi}}\right)
\simeq {\rm max}\left[
|a_{H}|\left(\frac{|\phi|_{0}}{M_{*}}\right)^2\delta_{\rm eff}^{H},\;\;
|a_{m}|\left(\frac{m_{3/2}}{m_{\phi}}\right)^2\left(
\frac{|\phi|_{0}}{M_{*}}\right)^2\delta_{\rm eff}^{\tilde{G}}
\right]\;,
\label{B-phi-ratio}
\end{equation}
where $\delta_{\rm eff}^{H}\equiv {\rm sin}({\rm arg}a_{H}+4{\rm
  arg}(\phi_{0}))$, and $\delta_{\rm eff}^{\tilde{G}}\equiv {\rm
  sin}({\rm arg}a_{m}+ 4{\rm arg}\phi_{0})$.~\footnote{This ratio
  cannot exceed $1/6$ for our choice of the flat direction. If
  $m_{3/2}\gg m_{\phi}$, as in the case of anomaly-mediated
  SUSY-breaking models, the condition $|\phi|_{0}\lsim
  m_{\phi}M_{*}/m_{3/2}$ is necessary for the $\phi$ field not to be
  trapped at the local (or global) minimum located near the Planck
  scale~\cite{WKY}.}  Note that this ratio remains constant until the
$\phi$ field eventually decays into the SM particles and the LSPs.

After dozens of oscillations, the scalar condensate of the
$\phi$ field fragments into the Q-balls. Almost all the baryonic charge and
energy density carried by the coherent 
$\phi$ field are absorbed into the Q-balls~\cite{KK-1,KK-2}.
As for the details about the Q-ball, see Ref.~\cite{higgsino-wino2}.
The produced Q-balls behave as ordinary matter, and their energy density
decreases as $\rho_{Q}(=\rho_{\phi})\propto R^{-3}$, where $R$ is the scale factor of the
expanding Universe.  On the other hand, the energy of the
inflaton is converted into  radiation through reheating. After
completion of the reheating process, the energy density of the radiation
decreases as $\rho_{R}\propto R^{-4}$.
Then, the energy density of the Universe is dominated by the
produced Q-balls before their decays, which take place at $T=T_{d}$, 
when the following condition is satisfied:
\begin{equation}
 T_{R}>3 T_{d}\left(\frac{M_{*}}{|\phi|_{0}}\right)^2\;,
\label{condition-for-domination}
\end{equation}
where $T_{R}$ is the reheating temperature of the inflation.
Although it depends on the size of the Q-ball charge, the typical decay
temperature of the Q-ball lies in the range $10\MEV\lsim T_{d}\lsim
(\mbox{a few})\GEV$, and hence 
the relation Eq.~(\ref{condition-for-domination})
can be easily satisfied. In the following,
we assume this to be the case. We will come back to this
condition after the derivation of Eq.~(\ref{therelation}).

The resultant baryon asymmetry after the decays of the
Q-balls is now given by the following simple formula:
\begin{eqnarray}
 \frac{n_{B}}{s}&=&\frac{\rho_{Q}}{s}\left(\frac{n_{\phi}}{\rho_{Q}}\right)
\left(\frac{n_{B}}{n_{\phi}}\right)=\frac{3}{4}\frac{T_{d}}{m_{\phi}}
\left(\frac{n_{B}}{n_{\phi}}\right)\nonumber\;,
\label{baryon-asymmetry}
\end{eqnarray}
where $s$ is the entropy density of the Universe.
Here, in the second equality, we have used the fact that the effective mass of
the Q-ball per $\phi$-number is given by $m_{\phi}$ in a good
approximation.
Note that the resultant baryon asymmetry is completely independent of
the detailed history of the Universe, since
the Q-balls dominate the energy density  of the Universe 
before they decay.

If the LSP $\chi$ has a large $s$-wave annihilation cross 
section, 
which is the case for the $\widetilde{H}$- or
$\widetilde{W}$-like LSP,~\footnote{The sneutrino $\tilde{\nu}$ is also an
interesting candidate. Even the bino-like LSP can have a large $s$-wave
component in the case of very large ${\rm tan}\beta$ via 
$\widetilde{H}$ contamination.} the resultant relic abundance of $\chi$ 
is given by
the following simple expression to quite good
accuracy~\cite{higgsino-wino1,higgsino-wino2}:
\begin{equation}
 \frac{n_{\chi}}{s}=\sqrt{\frac{45}{8\pi^2 g_{*}(T_{d})}}\;
\frac{\vev{\sigma v}_{\chi}^{-1}}{M_{*}T_{d}}\;,
\label{neutralino-number}
\end{equation}
where $n_{\chi}$ is the number density of the LSP, and
$g_{*}(T_{d})$ denotes the relativistic degrees of freedom at
$T=T_{d}$.
In terms of the density parameter, the resultant matter density of baryons
and that of dark matter can be expressed as follows:
\begin{eqnarray}
 &&\Omega_{B}h^2\approx 0.02\left(\frac{1\TEV}{m_{\phi}}\right)
\left(\frac{T_{d}}{100\MEV}\right)\left[\left(
\begin{array}{ccc}
|a_{H}|\left(\displaystyle{\frac{|\phi|_{0}}{M_{*}}}\right)^2\delta_{\rm eff}^H\\
{\rm or}\\
|a_{m}|\left(\displaystyle{\frac{m_{3/2}}{m_{\phi}}}\right)^2
\left(\displaystyle{\frac{|\phi|_{0}}{M_{*}}}\right)^2\delta_{\rm eff}^{\tilde{G}}
\end{array}\right)\times 10^{6}
\right],
\label{baryon}\\\nonumber\\
&&\Omega_{\chi}h^2\approx 0.3 \left(\frac{10}{g_{*}(T_{d})}\right)^{1/2}
\left(\frac{m_{\chi}}{100\GEV}\right)\left(\frac{100\MEV}{T_{d}}\right)
\left(\frac{10^{-7}\GEV^{-2}}{\vev{\sigma v}_{\chi}}\right).
\label{dark}
\end{eqnarray}
Therefore, if either $\widetilde{H}$ or $\widetilde{W}$ is the LSP,
the baryon asymmetry and the dark matter may be 
explained simultaneously
in the following range of parameters:
\begin{eqnarray}
&&10^{15}\GEV\lsim |\phi|_{0}\lsim 10^{17}\GEV\;\; ,
\;\;10\MEV\lsim
T_{d}\lsim (\mbox{a few})\GEV\;,\nonumber\\
&&10^{-8}\GEV^{-2}\lsim \vev{\sigma v}_{\chi}\lsim 10^{-7}\GEV^{-2}\;\;,
\;\; 10^{-2}\lsim |a|\delta_{\rm eff}\lsim 10^{-1}\;,
\end{eqnarray}
where $a$ and $\delta_{\rm eff}$ denote either $a_{H}$ or $a_{m}$,
and either $\delta_{\rm eff}^{H}$ or $\delta_{\rm eff}^{\tilde{G}}$,
respectively.
\section{Derivation of the relation of $\Omega_{B}$ and $\Omega_{DM}$}
Now, we come to deriving the relation in  
Eq.~(\ref{therelation}).
For this purpose, we have to know the $|\phi|_{0}$ dependence on the
decay temperature $T_{d}$ of the Q-ball. The terms $|\phi|_{0}$ and $T_{d}$
are related to each other through
the size of the Q-ball charge ``$Q$''.

The size of the Q-ball charge crucially depends on the initial amplitude
of the $\phi$ field and its scalar potential. The scalar potential
relevant at the time of the Q-ball formation can be written as
\begin{equation}
V(\phi)=m_{\phi}^2\left(1+K{\rm log}\left(
\frac{|\phi|^2}{M_{G}^2}
\right)\right)|\phi|^2\;,
\label{potetial-q-ball-formation}
\end{equation}
where $M_{G}$ is the renormalization scale at which the
soft mass $m_{\phi}$ is defined, and the $K{\rm log}(|\phi|^2)$
term represents the one-loop correction.
This mainly comes from the gluino loops and the typical value of $K$
is found to be in the range $-0.1\lsim K\lsim -0.01$~\cite{K-factor}.
This negativeness of the $K$-factor makes the potential a little bit
flatter than the quadratic one, which is a necessary and sufficient  condition for the
Q-ball to be formed in the present baryogenesis.

Recently, the typical size of the Q-ball charge has been calculated by
detailed lattice simulations~\cite{KK-1,KK-2}. Applying their result to our model, the
size of the Q-ball charge can be estimated to
\begin{equation}
 Q=\bar{\beta}\left(\frac{|\phi|_{0}}{m_{\phi}}\right)^2 \epsilon\;,
\label{q-ball-charge}
\end{equation}
where $\bar{\beta}\simeq 6\times 10^{-3}$, which depends on the
$K$-factor and the fluctuations of the $\phi$ field after  inflation, 
and $\epsilon\simeq 0.01$ is a constant independent of
$|\phi|_{0}$.~\footnote{This is the case for
$(n_{B}/n_{\phi})
\lsim 0.01$, which is naturally satisfied in the present model with
$|\phi|_{0}\lsim M_{*}$. If this is not the case,
$\epsilon$ should be replaced
by $(n_{B}/n_{\phi})$.
}

The size of the charge that evaporates from the surface of a single
Q-ball through interactions with thermal backgrounds is about $\Delta
Q\sim 10^{18}$~\cite{diffusion}.  Therefore, as long as
$|\phi|_{0}\gsim 10^{14}\GEV$, the Q-ball survives thermal
evaporation. The remaining charges of the Q-ball are emitted through
its decay into light fermions. The decay rate was calculated in
Ref.~\cite{Q-decay} as
\begin{equation}
 \Gamma_{Q}\equiv -\frac{dQ}{dt}\lsim \frac{\omega^3 {\cal A}}{192\pi^2}\;.
\label{decay-rate}
\end{equation}
Here, ${\cal A}=4 \pi R_{Q}^2$ is the surface area of the Q-ball,
where $R_{Q}\simeq \sqrt{2}/(m_{\phi}\sqrt{|K|})$ is the Q-ball
radius;  $\omega\simeq m_{\phi}$ is the effective mass of the Q-ball
per $\phi$-number. This upper bound is likely to be saturated for
$\phi(0)\gg m_{\phi}$, where $\phi(0)$ is the field value of 
$\phi$ at the centre of the Q-ball. This is the case in the present model.

Then, the decay temperature of the Q-ball is given by
\begin{eqnarray}
 T_{d}&=& \frac{\eta}{\sqrt{48 |K|\pi}}\left(\frac{90}{\pi^2 g_{*}(T_{d})}\right)^{1/4}
\left(\frac{m_{\phi}M_{*}}{Q}\right)^{1/2}\;,
\label{decay-temperature}\\
&\simeq& 2\GEV\times \eta\left(\frac{0.03}{|K|}\right)^{1/2}\left(
\frac{m_{\phi}}{1\TEV}\right)^{1/2}\left(\frac{10^{20}}{Q}\right)^{1/2}\;,
\label{decay-temp-nume}
\end{eqnarray}
where $\eta\lsim 1$ denotes the ambiguity coming from an inequality of
the decay rate in Eq.~(\ref{decay-rate}).~\footnote{This factor is, in
principle, numerically calculable by determining the accurate profile of
the Q-ball, which requires the details of the scalar potential.
There might also exist an ambiguity from the decay channels induced by loop
diagrams.
}
Then, in terms of the initial amplitude of the $\phi$ field, the
decay temperature is written as follows (see Eq.~(\ref{q-ball-charge})):
\begin{equation}
 \frac{T_{d}}{m_{\phi}}=\frac{\eta}{\sqrt{48\pi|K|\bar{\beta}\epsilon}}\left(\frac{90}{\pi^2 g_{*}(T_{d})}\right)^{1/4}\left(\frac{m_{\phi}M_{*}}{|\phi|_{0}^{2}}\right)^{1/2}\;.
\label{decay-temp-2}
\end{equation}

From Eqs.~(\ref{baryon}), (\ref{dark}) and (\ref{decay-temp-2}), we can
see that both the matter density of baryons and that of dark matter are 
linearly proportional to $|\phi|_{0}$, and hence 
we can easily derive the wanted
relation:
\begin{equation}
 \frac{\Omega_{B}}{\Omega_{\chi}}=
\frac{\eta^2}{16\pi|K|\bar{\beta}\epsilon}\left(\frac{m_{p}}{m_{\chi}}\right)
\left(\frac{m_{\phi}^2}{\vev{\sigma v}_{\chi}^{-1}}\right)
\times{\rm max}\left\{
\begin{array}{cc}
|a_{H}|\delta_{\rm eff}^{H}\\
\\
|a_{m}|\left(\displaystyle{\frac{m_{3/2}}{m_{\phi}}}\right)^2\delta_{\rm eff}^{\tilde{G}}
\end{array}
\right\}\;.
\label{analytic-relation}
\end{equation}
If we assume $(|a_{H}|,\;|a_{m}|,\;\eta)\approx1$ and the typical values
for the parameters, $K$, $\bar{\beta}$ and $\epsilon$,
the above  equation is
written as
\begin{equation}
 \frac{\Omega_{B}}{\Omega_{\chi}}\approx
10^{3\mbox{--}4}\left(\frac{m_{p}}{m_{\chi}}\right)
\times {\rm max}\left[\left(
\frac{m_{\phi}^2}{\vev{\sigma v}_{\chi}^{-1}}
\right)\delta_{\rm eff}^{H}\;\;,\;\;
\left(\frac{m_{3/2}^2}{\vev{\sigma v}_{\chi}^{-1}}\right)\delta_{\rm eff}^{\tilde{G}}
\right]\;.
\label{therelation2}
\end{equation}
By using the typical annihilation cross section for $\widetilde{H}$- or
$\widetilde{W}$-like LSP, $\vev{\sigma v}_{\chi}\simeq 5\times
10^{-8}\GEV^{-2}$, we derive
\begin{equation}
 \frac{\Omega_{B}}{\Omega_{\chi}}\approx 0.15\times \left(\frac{100\GEV}{m_{\chi}}\right)
\left(\frac{m}{1\TEV}\right)^2\left(\frac{\vev{\sigma v}_{\chi}}{5\times 10^{-8}\GEV^{-2}}\right)\left(\frac{\delta_{\rm eff}}{0.1}\right)\;,
\label{final-relation}
\end{equation}
where $m$ and $\delta_{\rm eff}$ denote either $m_{\phi}$ or $m_{3/2}$,
and either $\delta_{\rm eff}^{H}$ or $\delta_{\rm eff}^{\tilde{G}}$,
respectively. This is perfectly consistent with observations.

We can obtain another interesting piece of 
information by ``multiplying'' the mass
density of baryons and that of dark matter together.
In this case, we can remove the ambiguity associated with the decay
temperature of the Q-ball.
It is easy to show that the following relation holds:
\begin{equation}
\Omega_{B}h^2\times\Omega_{\chi}h^2=1.76\times 10^{-2}\;
\frac{m_{\chi}}{m_{\phi}}\left(\frac{\vev{\sigma v}_{\chi}^{-1}}{{\rm GeV}^{2}}\right)
\frac{1}{\sqrt{g_{*}(T_{d})}}\left(\frac{n_{B}}{n_{\phi}}\right)\;.
\end{equation}
This relation gives us information on $|\phi|_{0}$, the initial 
amplitude of the $\phi$ field.
If we take the typical values for parameters,
we can see that
the present mass density of baryons and that of dark matter
suggest 
$10^{15}\GEV\lsim |\phi|_{0}\lsim 10^{16}\GEV$,
which surprisingly coincides with  the $B-L$ breaking scale suggested from
the see-saw neutrino masses~\cite{GRSY}.
Note that this value does not affect the prediction in
Eq.~(\ref{analytic-relation}).

\section{Conditions for the relation to hold}
In this section, we discuss the conditions for relation 
(\ref{analytic-relation}) to hold.
We have to examine the thermal effects, on the scalar potential, of the
$\phi$ field given in Eq.~(\ref{potential-phi}).
If the thermal effects dominate the scalar potential when the $\phi$
field starts to  oscillate,
they cause the evaporation of the $\phi$ field before the
formation of the Q-balls, or drastically change the $|\phi|_{0}$
dependence on the Q-ball charge.

First, let us discuss the effects related with the thermal mass terms.
If the cosmic temperature $T$ satisfies $f|\phi|_{0}<T$, the field
coupled to the $\phi$ field through the coupling constant $f$ induces
the thermal mass term $c^2f^2T^2|\phi|^2$, where $c$ is a real
constant of the order of $1$.  Therefore, if the two conditions,
$f|\phi|_{0}<T$ and $cfT>H$, are satisfied simultaneously when $H\gsim
m_{\phi}$, the thermal mass term causes the early oscillation of the
$\phi$ field~\cite{DRT,earlyosc,earlyosc-sys}.  In this case, the $\phi$
field is thermalized and Q-balls are not formed.  The condition for
the reheating temperature of inflation, $T_{R}$, to avoid this early
oscillation is given by~\footnote{Here, we have used the fact that the cosmic
  temperature behaves as $T\simeq (H T_{R}^2M_{*})^{1/4}$ before the
  completion of the reheating process of inflation.  }
\begin{equation}
 T_{R}\lsim {\rm max}\left\{
\frac{f}{\sqrt{c}}|\phi|_{0}\left(\frac{|\phi|_{0}}{M_{*}}\right)^{1/2}\;,\;
\frac{m_{\phi}}{c^2 f^2}\left(\frac{m_{\phi}}{M_{*}}\right)^{1/2}
\right\}
\label{T-condition2}
\end{equation}
This condition should be satisfied in  all the gauge and Yukawa coupling
constants associated with the $\phi$ field.  The most stringent
constraint comes from the first term with Yukawa coupling of an up quark,
$Y_{u}\simeq 10^{-5}$.  This constraint becomes significantly weak if
we adopt the second and third generations of squarks for the flat direction.

A much more stringent constraint comes from the thermal logarithmic 
potential~\cite{thermal-log}:
\begin{equation}
 \delta V\supset a_{g} \alpha\; T^4{\rm log}\left(
\frac{|\phi|_{0}^{2}}{M_{*}^2}\right)\;,
\label{thermal-log}
\end{equation}
where $|a_{g}|={\cal{O}}(1)$, and $\alpha$ is a constant
given by the fourth power of gauge and/or Yukawa coupling constants.
This leads to the following constraint on the reheating 
temperature~\cite{ADwithBL,R-independent}:
\begin{equation}
 T_{R}\lsim \frac{1}{\sqrt{|a_{g}|\alpha}}\left(\frac{m_{\phi}}{M_{*}}\right)^{1/2}
|\phi|_{0}\;.
\label{T-condition3}
\end{equation}

\begin{figure}[h!]
 \centerline{\psfig{figure=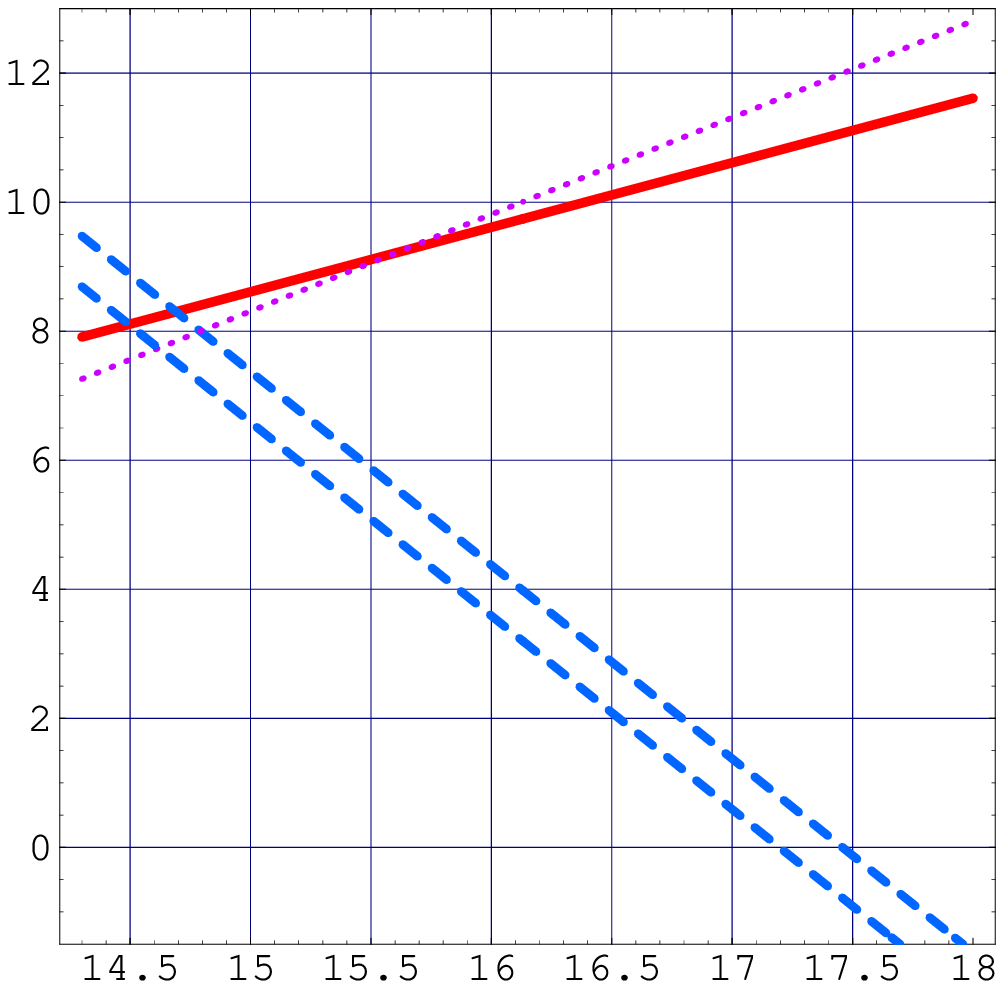,height=7cm}}
 \begin{picture}(0,0)
\small
  \put(65,110){${\rm log}_{10}\left[\displaystyle{\frac{T_{R}}{{\rm GeV}}}\right]$}
  \put(200,-5){${\rm log}_{10}\left[\displaystyle{
\frac{|\phi|_{0}}{\rm GeV}}\right]$ }
\normalsize
 \end{picture}
\vspace{0.1cm}
 \caption{The allowed region of reheating temperature for 
   relation (\ref{analytic-relation}) to hold.  The red (solid)
   line denotes the upper bound from Eq.~(\ref{T-condition3}) with
   $|a_{g}|=1$ and $\sqrt{\alpha}=1/20$. The purple (dotted) line
   denotes the upper bound from the thermal mass term,
   Eq.~(\ref{T-condition2}). Here, we take $c=1$, $f=10^{-5}$. This
   constraint becomes significantly weak if we do not adopt the first
   generation of up-type squark in the flat direction.  The blue
   (dashed) lines represent the lower bounds from the Q-ball dominance
   condition given in Eq.~(\ref{condition-for-domination}) with
   $\eta=0.3,\;|K|=0.1$ and $\eta=1\;,|K|=0.03$ from the bottom up,
   respectively. Throughout this calculation, we have used
   $m_{\phi}=1\TEV$.  }
 \label{FIG-TR}
\end{figure}

From Eqs.~(\ref{T-condition2}) and (\ref{T-condition3}) combined with
the Q-ball dominance condition in
Eq.~(\ref{condition-for-domination}), we obtain the allowed region of
the reheating temperature where the relation
Eq.~(\ref{analytic-relation}) holds.  We show this region in
Fig.~\ref{FIG-TR}.

From this figure, we see that there is a  wide allowed region for
the reheating temperature if $|\phi|_{0}\gsim 10^{15}\GEV$.  Note that,
even if we use $T_{R}\gg10^{8}\GEV$, there is no ``cosmological
gravitino problem''~\cite{gravitino} 
as emphasized in Ref.~\cite{higgsino-wino2}. This is
because the large entropy production associated with the decays of Q-balls
dilutes the gravitino number density substantially.
As long as the reheating temperature and the initial amplitude of the
$\phi$ field are within this allowed region,
$\Omega_{B}/\Omega_{\chi}$ is  fixed by the low-energy
parameters, and is totally independent of $T_{R}$ and $|\phi|_{0}$.

\section{Conclusions and discussion}
In this paper, we have proposed a solution to the
$\Omega_{B}$--$\Omega_{DM}$ coincidence puzzle.  We have pointed out
that a class of the AD baryogenesis directly relates the observed mass
density of baryons to that of dark matter with the help of the
late-time decays of Q-balls, if the LSP has a large $s$-wave
annihilation cross section, as in the case of $\widetilde{H}$- or
$\widetilde{W}$-like LSPs.  The relation we have shown is~\footnote{In
  the case of $m_{3/2}>m_{\phi}$, we should replace $m_{\phi}$ by
  $m_{3/2}$.}
\begin{equation}
 \frac{\Omega_{B}}{\Omega_{\chi}}\approx 10^{3{\mbox{--}}4}\left(
\frac{m_{\phi}^2}{\vev{\sigma v}_{\chi}^{-1}}\right)
\left(\frac{m_{p}}{m_{\chi}}\right)\delta_{\rm eff}\;.
\label{therelation-conclusion}
\end{equation}

A beautiful point of this scenario
 is that the ratio is totally independent of the
reheating temperature of inflation and the initial amplitude of the
$\phi$ field, as well as from a  detailed history of the Universe, once
the relatively weak constraints shown in Fig.~\ref{FIG-TR} are satisfied.  
We should stress that the ratio is solely determined by the low-energy
parameters, except for the ${\cal O}(0.1)$ effective CP phase, which
makes this scenario quite testable in the future experiments.  Our
scenario is  free from the ``cosmological gravitino problem'', since
the large entropy production associated with the Q-ball decay
sufficiently dilutes the number density of
gravitinos.

In addition to the above beautiful relation, there are many interesting
implications of the present scenario.
The large $\widetilde{H}$ or $\widetilde{W}$ component of the LSP
significantly enhances its detection rates in both direct and indirect
dark matter searches~\cite{higgsino-wino2}. 
Furthermore, such non-thermal dark matter may naturally 
explain the observed excess of positron flux in cosmic 
rays~\cite{positron}.

\section*{Acknowledgements}
M.F. thanks the Japan Society for the Promotion of Science for
financial support.  This work was partially supported by Grant-in-Aid
for Scientific Research (S) 14102004 (T.Y.).

\small


\end{document}